\documentclass[runningheads]{llncs}
\usepackage{graphicx}
\usepackage[ruled,vlined]{algorithm2e}

\usepackage{url}
\usepackage{xcolor}

\usepackage{hyperref}

\usepackage{amsmath}
\usepackage{amsfonts}
\usepackage{multirow}
\usepackage{multicol}

\begin{document}

\title{ Histogram of Oriented Gradients Meet Deep Learning: A Novel Multi-task Deep Network for Medical Image Semantic Segmentation}%

\author{Binod Bhattarai\inst{1} \and
Ronast Subedi$^*$\inst{2} \and
Rebati Raman Gaire$^*$\inst{2}  \and
Eduard Vazquez\inst{3} \and 
Danail Stoyanov \inst{1}} 
\authorrunning{Bhattarai et al.}
\institute{
University College London, UK\\
\and
NAAMII, Nepal \and
Redev Technology, UK\\
% \url{http://www.springer.com/gp/computer-science/lncs} 
\email{\{b.bhattarai,danail.stoyanov\}@ucl.ac.uk}}

\maketitle

\def\thefootnote{*}\footnotetext{Equal contributions}

\begin{abstract}
We present our novel deep multi-task learning method for medical image segmentation. Existing multi-task 
methods demand ground truth annotations for both the primary and auxiliary tasks. Contrary to it, we propose to
generate the pseudo-labels of an auxiliary task in an unsupervised manner. To generate the pseudo-labels, we
leverage Histogram of Oriented Gradients (HOGs), one of the most widely used and powerful hand-crafted 
features for detection. Together with the ground truth semantic segmentation masks for the primary task and 
pseudo-labels for the auxiliary task, we learn the parameters of the deep network to minimise the loss of both 
the primary task and the auxiliary task jointly. 
We employed our method on two powerful and widely used semantic segmentation networks: UNet and U2Net to 
train in a multi-task setup. To validate our hypothesis, we performed experiments on two different medical 
image segmentation data sets. From the extensive quantitative  and qualitative results, we observe that 
our method consistently improves the performance compared to the counter-part method. Moreover, our method
is the winner of FetReg Endovis Sub-challenge on Semantic Segmentation organised in conjunction with MICCAI 2021. 
For the code and implementation details, please \href{https://github.com/thetna/medical_image_segmentation}{click here}.

\end{abstract}
\section{Introduction}
\label{sec:intro}

\begin{figure*}
    \centering
    \includegraphics[width=1.0\linewidth]{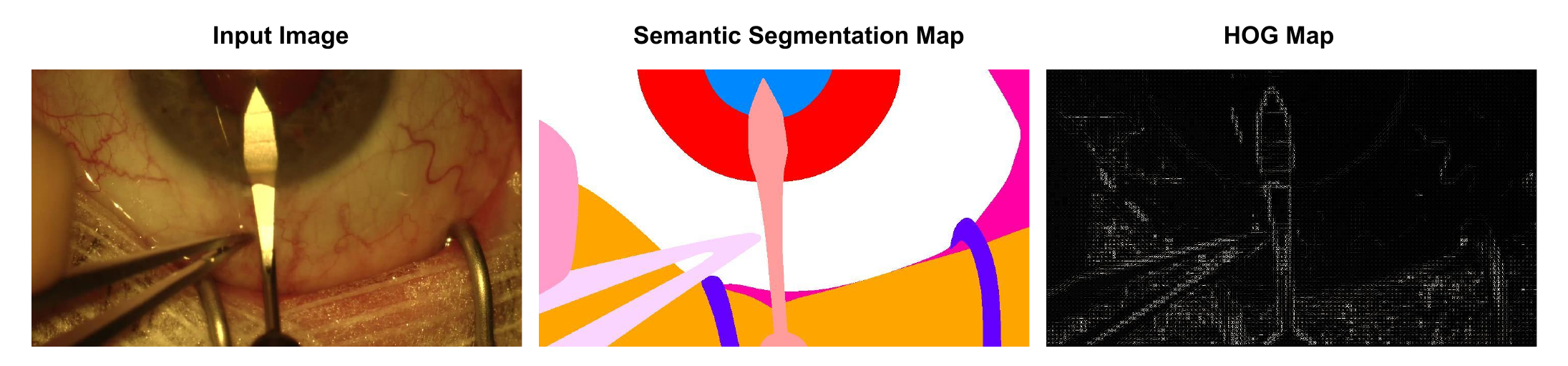}
    \caption{ This Figure shows an input image (left) and its ground truth semantic segmentation map (left) for the primary task and the Histogram of Oriented Gradients of the input image (right). 
    In the HOG map, we can observe the boundary between the organs and the instruments that belong to different semantic categories. Zoom in for 
    a better view.}
    \label{fig:hog_motivation}
\end{figure*}

Medical image segmentation~\cite{lei2020medical,milletari2016v,sharma2010automated,pham2000current} is an important and active research problem. The usage of semantic segmentation in 
several biomedical applications such as computer-assisted diagnosis~\cite{zhao2019computer}, 
robotic surgery~\cite{colleoni2020synthetic}, radiotherapy planning and follow-ups~\cite{nemoto2020simple}, etc., 
is growing day by day. Due to this reason, the research community has witnessed an 
unprecedented growth of research interest in this domain. There are several types of semantic segmentation 
problems in medical imaging. Broadly, the existing semantic segmentation tasks can be grouped into four 
major categories viz. organ segmentation~\cite{hu2017automatic}, robotic-instrument segmentation~\cite{pakhomov2019deep,shvets2018automatic}, vessels segmentation~\cite{fraz2012blood}, and cellar and sub-cellular segmentation~\cite{rizk2014segmentation}, etc.

After the seminal work of ~\cite{krizhevsky2012imagenet} on large-scale image classification using deep convolutional neural networks,
the use of deep architectures has not been limited only on computer vision~\cite{VGG_2015_ICLR,Googlenet_cvpr2015,resnet_cvpr2016}; 
it is equally popular in medical image analysis~\cite{suzuki2017overview,lee2017deep}. With the usage of deep learning algorithms, the accuracy of computer vision tasks such as classification, segmentation, and detection is improving significantly~\cite{rawat2017deep}. A similar trend has been observed on medical image analysis too~\cite{anwar2018medical}. We obtain the performance gain at the cost of many annotated examples (e.g. Imagenet consists of 1M annotated examples). It is evident that deep learning algorithms are data voracious and demand millions of training examples. Collecting data, in general, is time-consuming, needs experts and is also expensive. Moreover, in medical imaging, it is not only about collecting annotations as they come from highly trained experts, e.g. radiologists (e.g., MRI or CT scanner), but due to growing concerns on privacy, it is difficult to get the unlabelled examples
~\cite{peng2021self}. 

To improve the generalisation of a model from a fixed amount of training examples, sharing the parameters between main task 
and auxiliary tasks~\cite{caruana1997multitask} is popular for a long time. MaskRCNN~\cite{he2017mask} , one of the most popular networks in recent time, shares the parameters between detection and segmentation networks. Similarly,~\cite{takikawa2019gated} proposed to predict 
contour as an auxiliary task while training a network for semantic segmentation as the primary task. The major 
drawback of these methods is a need of annotated examples for both the primary and the auxiliary tasks. Collecting 
such a heterogeneously labelled set of training examples is even more challenging in the medical image domain.

To tackle the problem of collecting training examples with the heterogeneous set of labels, we propose to generate pseudo-labels for the auxiliary task from the hand-crafted features instead. As one can extract hand-crafted features in an unsupervised manner, generating pseudo-labels of any type of images for an auxiliary task can be done easily. 
To this end, we leverage the Histogram of Oriented Gradients (HOGs)~\cite{dalal2005histograms} to generate pseudo-labels. 
Demarcation of the organs and surgical instruments parts belonging to a common category from unrelated ones would play a significant role in their accurate segmentation. Auxiliary tasks focusing on such aspects would help the network to learn the robust representation for semantic segmentation. Thus, we chose HOGs to generate pseudo-labels for the auxiliary task as these features are carefully designed state-of-the-art hand-crafted features for object detection ~\cite{dalal2005histograms}. 
However, any other type of hand-crafted features can be employed in our pipeline to extract the pseudo-labels. Figure~\ref{fig:hog_motivation} shows the HOGs map of eye anatomy and surgical instrument. In the Figure, we can see the demarcation of a surgical instrument from eye anatomies made by the map of the Histogram of Oriented Gradients. Once, we extract the HOGs, we consider these representations as annotations of the auxiliary task and the ground truth semantic map as annotations of the primary task.  We extended existing popular architectures for semantic segmentation: UNet~\cite{ronneberger2015u} and U2Net~\cite{qin2020u2} to minimise the loss of both the auxiliary and primary tasks and train the network in a multi-task manner.

Use of image feature representations as a pseudo-label is growing these days.
Recently, ~\cite{gidaris2020learning} trained a deep network to predict Bag of Visual Words (BoWs) for image classification. Unlike ours, this method relied on the learned features extracted from a network trained to minimize the image rotation angle loss. In medical imaging, organs such as the eye bulb, pupils, colons, etc.,  are either hollow and cylindrical or rotationally invariant. Hence, the pipeline is not directly applicable in medical imaging. In addition, they trained their method to minimise the objective function of a single task, whereas we train our pipeline in a multi-task set-up. We summerise our contributions in the following points:

\begin{itemize}
    \item We investigated the Histogram of Oriented Gradients to generate pseudo-labels of images and exploited these representations as labels of an auxiliary task. 
    \item We extended existing semantic segmentation networks to train in a multi-task framework.
    \item We applied our method on two challenging medical semantic segmentation data sets. Our extensive experiments demonstrate that our pipeline consistently outperforms the counter-part single task networks. 
\end{itemize}    

\section{Related Works}
\label{related_works}
Our work falls into the category of deep multi-task learning with pseudo labels, self-supervised learning. In this Section, we summarise some of the important past works closely related to our method. 

\noindent \textbf{Deep Multi-task Learning for Semantic Segmentation:} 
UNet~\cite{ronneberger2015u} is one of the earliest and the most widely used deep architectures for medical image segmentation. This architecture is a supervised learning architecture and can handle only semantic maps as the ground-truth annotations. Another work on pancreas segmentation ~\cite{roth2018spatial} trains deep learning architecture in a multi-stage manner. It predicts the bounding box to localise the pancreas followed by fine-tuned semantic segmentation. Unlike our approach, this method uses ground truth annotations on both stages.
In contrast, we rely on HOGs computed unsupervised and trained the model to minimise the losses jointly.  Another work on brain lesion segmentation~\cite{kamnitsas2017efficient} employs 3D Convolutional Neural Network with a fully connected Conditional Random Field. Similarly, ~\cite{lei2020self} employ self co-attention to improve the performance of anatomy segmentation in whole breast ultrasound. 
However, these methods consider only semantic segmentation maps for ground truth.  One of the recent works on tumours segmentation in 3D breast ultrasound images~\cite{zhou2021multi} proposed to train CNN in multitasking fashion. ~\cite{wang2018simultaneous} modified UNet architecture to jointly minimise the segmentation and classification loss in ultra-sound images. ~\cite{xie2018breast} trained multi-stage multitask learning framework for breast tumour segmentation in ultrasound images.~\cite{song2020end} learns the parameters of network to minimize 
the loss for skin lesion detection, classification, and segmentation. ~\cite{chakravarty2018deep} trained a multi-task learning CNN for semantic segmentation and image level glaucoma classification. Another work on histopathology  image analysis~\cite{qu2019joint} trained a multi-task network for nucleus classification
and segmentation. All of these methods need ground truth annotations for both the main task (semantic segmentation) and auxiliary tasks. Whereas, in our case, we have annotations for the primary task and generate pseudo-labels for the auxiliary task.

\noindent \textbf{Self-supervised Learning:} 
In Self-supervised learning, the annotations for the pre-text tasks are generated in 
an unsupervised manner. In general, the parameters of a CNN are learned to minimise the loss of 
pre-text tasks followed by fine-tuning of the parameters for the downstream tasks. Several different ways 
are investigated in the past years to generate the annotations of pre-text tasks. These includes, 
image rotation angle~\cite{gidaris2018unsupervised}, colorization~\cite{zhang2016colorful}, image-patch 
context~\cite{pathak2016context}, in-painting~\cite{pathak2016context}, etc.  
These methods mostly pivot on the geometric transformations of the images. What kind of pre-text task is going
to be the most useful for the end-task is still an open research problem. 
Recently,~\cite{gidaris2020learning} proposed to learn the representations by predicting the visual Bag of Words (BoW). This method, closest to ours, rely on visual features to generate the pseudo-labels.
As we mentioned before, they compute BoWs from the visual representations extracted from model trained to minimise the rotation angle of an image. Thus, this approach is not directly 
applicable to our applications as most of the organs such as eyes, eye-bulb exhibit  rotationally invariant shape. 
Unlike most of the self-supervised pipeline, we propose to minimise the loss of end-task and
pre-text task jointly.

\section{Proposed Method}
\label{sec:proposed_method}
In this Section, we present our pipeline in detail. We start with the description of HOGs followed by the generation of pseudo-labels for the auxiliary task. Afterwards, we explain our approach to extend a 
single-task semantic segmentation network to a multi-task network. Finally, we explain the overall objectives.

We have a scenario $\mathcal{X} \times \mathcal{Y}$ where $\mathcal{X}$ represents input image space 
and $\mathcal{Y}$ represents output semantic map space. Our goal is to learn a function $f: \mathcal{X} \to \mathcal{Y}$ with a given training examples $T = \{(x_1, y_1), (x_2, y_2)\dots (x_i,y_i) \dots (x_N,y_N)\} \subset \mathcal{X} \times \mathcal{Y} $. In the training set $T$, $N$ is total number of training examples, $x_i \in \mathbb{R}^{(W \times H \times C)}$ , $y_i \in \mathbb{R}^{(W \times H)}$, where,  $W, H, C$ represents 
width, height, and total number of channels in an image respectively. Our contribution lies in generating 
extra annotations of the images in an unsupervised way and extending the single task semantic segmentation
network to train in a multi-task manner to improve the performance of semantic segmentation. We make use of HOGs to extract the pseudo-annotations of an image.

\subsection{Histogram of Oriented Gradients as Pseudo Labels}
\begin{figure*}
    \centering
    \includegraphics[trim= 0cm 0.5cm 0cm 0.0cm, clip, width=0.9\textwidth]{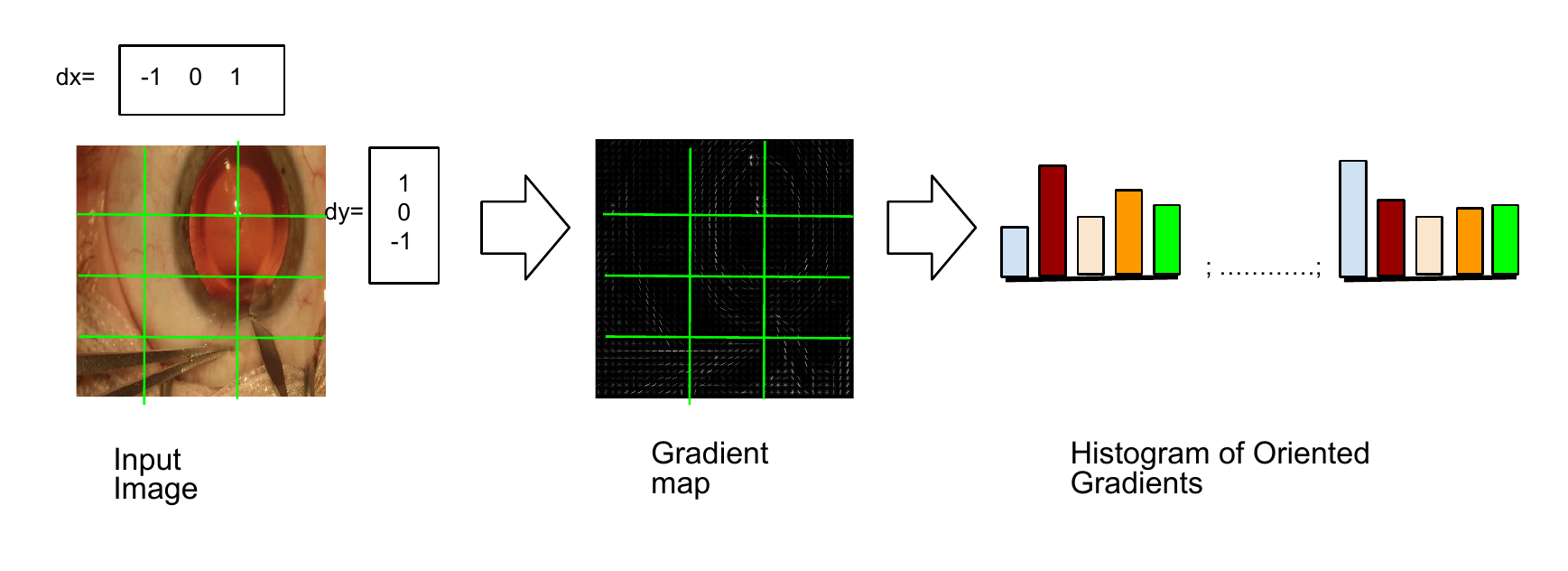}
    \caption{Diagram showing the pipeline to extract the Histogram of Oriented Gradients (HOGs). Zoom in for better view.}
    \label{fig:hog_pipeline}
\end{figure*}

It is proven that the HOGs~\cite{dalal2005histograms} were one of the most powerful hand-features on computer vision and medical image analysis especially for detection before the advent of data driven feature extraction methods 
such Alexnet~\cite{krizhevsky2012imagenet}, ResNet~\cite{resnet_cvpr2016}, and UNet~\cite{ronneberger2015u}.
In this paper, we use HOGs for a novel cause i.e. to extract the pseudo-labels of the images. 
% Figure~\ref{fig:hog_pipeline} shows our pipeline to extract the HOGs.
To compute HOGs from an image, first of all, we crop and resize the images to the desired dimensions of width, $W$ and height, $H$. We further divide the images into a non-overlapping
image patches of width $w$, and height $h$, resulting the total number of patches of $\lfloor W/w \rfloor \times  \lfloor H/h \rfloor$. For each of the patches, we run 1-D discrete derivative masks centred around a pixel in both the horizontal and vertical directions.  $d_x = [1, 0, -1]$ and $d_y= [1, 0, -1]^{T}$ are horizontal and vertical filtering kernels respectively. We run these filters on all the pixels of every image patches as shown in Figure~\ref{fig:hog_pipeline}.

\begin{figure*}
    \centering
    \includegraphics[trim=0cm 3.4cm 0cm 3.5cm, clip, width=0.99\linewidth] {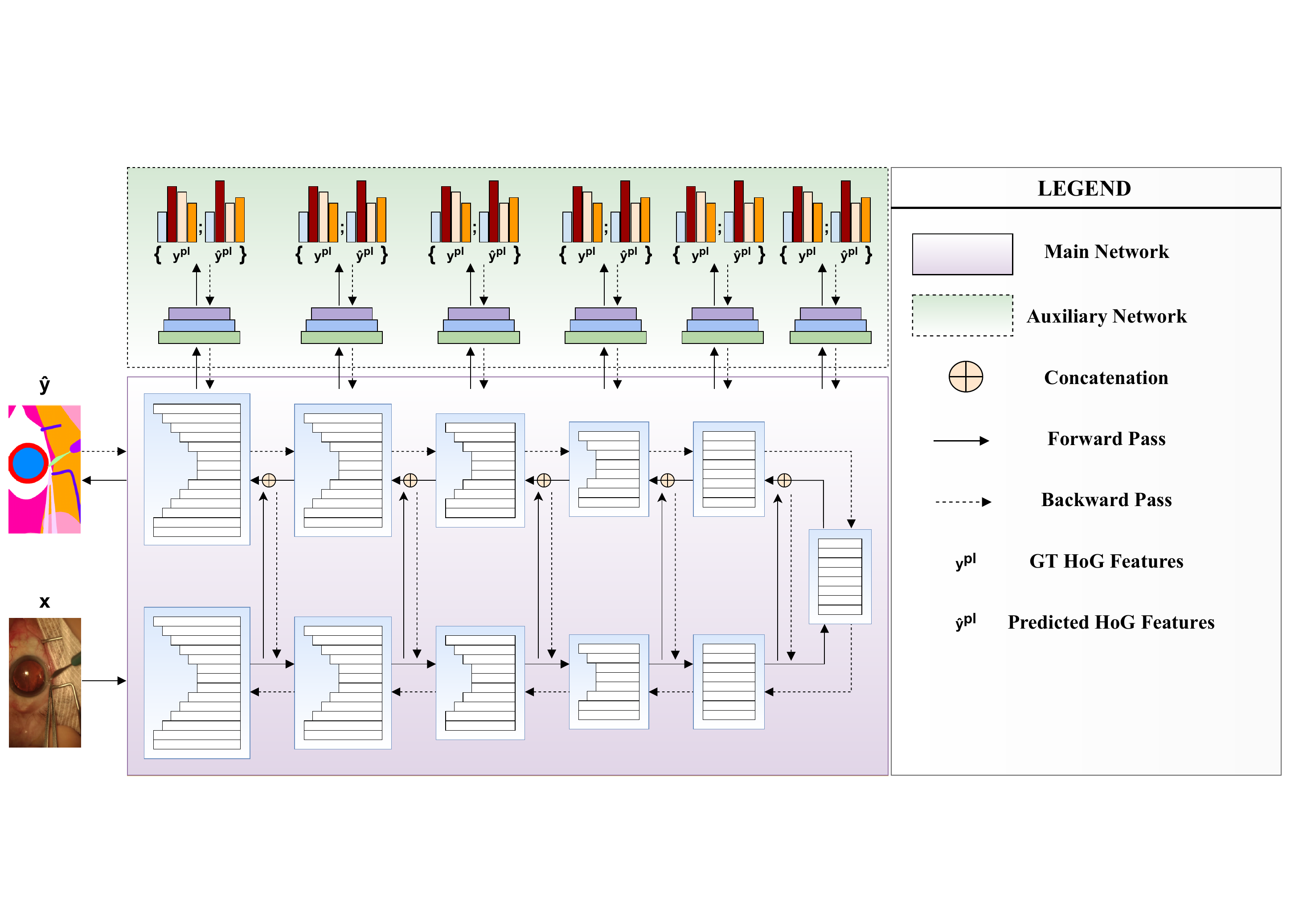}
    \caption{  
    This diagram shows the overall proposed framework. In the Figure, the main network corresponds to semantic segmentation network (e.g. U2Net), while the auxiliary network is our contribution to 
    extend the single task network to a multi-task network.  Training examples in triplet, i.e. input image, ground truth semantic map and pseudo-label computed from HOGs, are fed into the network and train the network jointly.
    }
    \label{fig:proposed_method}
\end{figure*}

After applying the kernels centred on every pixels, we compute the histogram of gradients for all the patches and append them together. Gradients are computed as $\arctan(\frac{d_y}{d_x})$, and the gradients are assigned to the nearest bin.  The histogram can have $k$ number of bins with angle ranging from 0 to 180 degrees. 
The magnitude of the gradient is computed as $\sqrt{d_x^2 + d_y^2}$. This magnitude of the gradients encodes the frequency of a bin of the gradient taken into consideration. In this manner, we estimate the histogram of oriented gradients in every patch.
The number of the bins and the patches determine the dimension of the HOGs and  are the hyper-parameters in our study. We present their studies in Experimental Section in depth. We concatenate the HOGs for all the patches of an image, and the final representations of  HOGs are the pseudo-label, $y^{pl}$ of the image. We augment the pseudo-label on the given training set. Thus, the training set with augmented pseudo-labels become $\{(x_i, y_i, y_i^{pl})\}_{i=1}^{i=N}$ which we use to train the semantic segmentation network in multi-task setup.

\subsection{Multi-task semantic segmentation with pseudo labels}
For an input image $x$ with the ground truth semantic segmentation map $y$ and 
its pseudo-label $y^{pl}$, we train a semantic segmentation network in a multi-task learning fashion. 
The primary task for us is to predict the semantic map and the secondary task is to regress the Histogram 
of Oriented Gradients (HOGs). To predict the semantic map we employ categorical cross-entropy loss and minimise mean squared loss to predict the HOGs. As mentioned before, UNet and U2Net are two most popular and the powerful semantic segmentation networks in medical imaging. However, these networks are originally designed to support semantic map
as only ground truth. Thus, these networks can not readily handle our heterogeneously labelled training examples. 
To enable them to handle pseudo-labels and share the parameters between these tasks, we proposed to add a regression unit with \textbf{two convolutional} layers and a fully connected layer on every layers of the decoder side on U2Net as shown in the Figure~\ref{fig:proposed_method}. On UNet, we added only one such unit on bottleneck. It is because, UNet has relatively less parameters compared to U2Net.  In the Figure~\ref{fig:proposed_method}, the left hand block depicts the U2Net architecture and the right hand side block shows the regression units we introduced in the architecture. The regression units learn the parameters predicts HOGs correctly. In the similar manner, we plugged in regression units on UNet. Compared to UNet, U2Net is also an hourglass architecture where each layer consists of a UNet. We learn the parameters of the the whole architecture to minimize the following objective. 

\begin{table}
\centering
\begin{tabular}{ c|c } 
\hline
%  \multicolumn{2}{c}{HOG Regressor} \\
% \hline
Input Shape & Operations \\
\hline
($3, h, h$) & Conv(3,3,1), ReLU(), MaxPool2d(2,2)  \\
\hline
(3, $\frac{h}{2}, \frac{h}{2}$) & Conv(3,3,1), ReLU(), MaxPool2d(2,2)  \\
\hline
(3, $\frac{h}{4}, \frac{h}{4}$) & Flatten()  \\
\hline
($3 \times \frac{h}{4} \times \frac{h}{4}$) & Linear(504) \\
% \hline
\end{tabular}
\label{tab:hog_network_architecture}
\caption{Architecture of the Auxiliary Task Network to Regress HOGs.}
\end{table}

\begin{equation}
   L = \frac{1}{N} \sum_{i=1}^{i=N}\alpha L_{ce} (x_i, y_i) + \beta L_{HOG}(x_i, y_i^{pl}) 
   \label{equation:overall_objective}
\end{equation}

In Equation~\ref{equation:overall_objective}, $L_{ce}$ is the primary task loss i'e minimization of cross-entropy loss to predict the ground truth mask correctly. Whereas, $L_{HOG}$ is loss of secondary task to predict the HOGs of the input image. We minimize the mean squared error between the predicted and ground truth HOGs. $\alpha$ and $\beta$ are two hyper-parameters to weight the contributions of each of the losses to best generalise the model parameters on unseen data for semantic segmentation. We fine-tune these parameters by doing cross-valiation on validation set.  The details are on Section~\ref{sec:experiments}.
\section{Experiments}
\label{sec:experiments}

\noindent \textbf{Data sets:} 
We evaluated our methods on two different publicly available challenging data sets with diverse characterstics. 
CaDIS data set ~\cite{grammatikopoulou2021cadis} was released in MICCAI 2020 in  
one of the EndoVis challenges. It consists of 25 surgical videos. Each video frame
is annotated broadly into eye anatomies, surgical instruments, and miscellaneous categories. Based on the granularity of the segments, ~\cite{grammatikopoulou2021cadis}  designed the challenge into three different tasks. Task 1 consists of 8 different segments: four eye anatomies, three misc objects, and one instrument category. 
In Task 2, the instrument category is further split into nine classes, resulting in 17 different categories. Finally, in Task 3,  there is an increase in granularity on the handles of the surgical instrument. This further increase in granularity resulted in 25 different categories to segment. There are 3,550 annotated frames in train set, 534 in validation set, and 586 are in test set. 

Another data set on which we evaluated our method is Robotic Instrument Segmentation~\cite{allan20192017}.  
This data set is publicly available for research since MICCAI 2017 challenge. The main task on this data set is to segment surgical instruments from the background. Based on the granularity of segmentation of the parts of the surgical instruments, three tasks were designed in the challenge. Task 1 is to segment
the instruments as a whole from the rest of the background. Similarly, the challenge in Task 2 is to segment the instruments into wrist, jaw, and shaft and distinguish the instrument from the background.  Finally, Task 3 further segments the instrument into seven parts and segregates it from the background.  There are  10 different folds of videos in total. Following the evaluation protocol described on ~\cite{allan20192017}, we report performance on folds 9 and 10 and train on rest of the videos.

\noindent \textbf{Baselines Architectures:} We took UNet~\cite{ronneberger2015u} and U2Net~\cite{qin2020u2}, two representative architectures, for semantic segmentation and employed our
method on these two architectures. Since our method is generic in nature, we can easily extend to other architectures.  UNet is one of the most widely used architectures in medical image segmentation. It is a lightweight architectures consisting of encoder and decoder. Encoder consists of convolutional and pooling  
layers that map high-dimensional images into low-dimensional latent space. Decoder feeds in the latent representations of the image and learns the parameters to predict the correct semantic maps. There are skip connections from encoder layers to decoder layers.

U2Net is another recently proposed architecture with state-of-the-art performance on multiple computer vision semantic segmentation benchmarks. Similar to UNet, this is an hourglass architecture with skip connections between the encoder and decoder layers. Compared to UNet, U2Net consists of UNet like structures in every layer of encoders and decoders and also known as  UNet inside UNet. Thus, the learning parameters in this architecture are much higher than UNet.

\noindent \textbf{Evaluation Metrics:} We used mean Intersection of Union (mIoU) to compare the quantitative performance.  Intersection of Union (IoU) is computed as follows:

\begin{align*}
   \text{IoU} = \frac{\text{true positive}}{\text{true positive + false positive + false negative}}
\end{align*}
% \text{Intersection of Union (IoU)} = \frac{true positive}{\text{true positive + false positive + false negative}}
% \]
In addition to this, we also present extensive qualitative analysis to make the comparisons. 

\noindent \textbf{Implementation Details:} 
 We implemented our algorithms on PyTorch framework. For optimization, we employ Adam Optimizer. We set the 
 initial learning rate to 2e-4 and scaled it by a factor of 0.5 in every 50k iteration. We train our 
 algorithms for 150K iterations and validate every 1k iterations. We save the best performing checkpoint 
 on the validation set and report the performance on the test set.

\begin{table*}[]
    \centering
    \begin{tabular}{c|c|c|c|c|c|c |c }
    \hline 
    \multirow{2}{*}{Task} & \multirow{2}{*}{ \# Classes} & \multicolumn{3}{c|}{Validation set mIOU} & \multicolumn{3}{l}{Test set mIOU} \\ 
     \cline{3-8}
      &  & MICCAI'21 & U2Net & +HOG (Ours) & MICCAI'21 & U2Net &  +HOG (Ours) \\
     \hline 
     1 & 8 & \textbf{86.7} & 84.9 & 85.5 & \textbf{83.7}  & 80.2 &  81.4 \\
     \hline
     2 & 18 & 72.7 & 83.8 & \textbf{84.1} & 70.6  & 77.8 & \textbf{80.2} \\
     \hline 
     3 & 26 & 66.6 & 82.1 & \textbf{83.0} & 59.2  & 78.2 & \textbf{78.4}   \\
     \hline 
     \end{tabular}
    \caption{Summary of quantitative performance comparison on CaDIS data set.}
    \label{tab:quant_cataract}
\end{table*}

\begin{table*}[]
    \centering
    \begin{tabular}{c|c|c|c|c|c|c |c }
    \hline 
    \multirow{2}{*}{Task} & \multirow{2}{*}{ \# Classes} & \multicolumn{3}{c|}{ mIOU on test Video 9}  & \multicolumn{3}{l}{mIOU on test Video 10} \\ 
     \cline{3-8}
      &  & MICCAI'17 & U2Net & +HOG (Ours) & MICCAI'17 & +HOG (Ours) &  Ours \\
     \hline
     1 & 2 & 87.7 & 94.2 & \textbf{95.6} & 91.7 & 96.0 & \textbf{96.2} \\
     \hline
     2 & 4 & 73.6 & 70.8 & \textbf{75.8} & 80.7 & 84.1 & \textbf{84.4} \\
     \hline 
     3 & 8 & 35.7 & 57.9 & \textbf{65.4} & 79.1 & 89.4 & \textbf{91.3} \\
     \hline 
     \end{tabular}
    \caption{Summary of quantitative performance comparison on Robotic Instrument Segmentation data set.}
    \label{tab:quant_robotic_instrument}
\end{table*}

\begin{figure}
    \centering
    \includegraphics[trim= 0cm 0.0cm 0cm 1.5cm, width=0.95\linewidth]{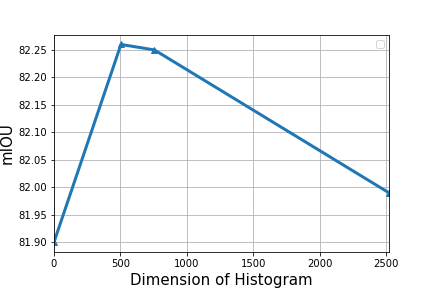}
    \caption{A performance comparison with varying dimensions of HOGs.}
    \label{fig:ablation_hog_dimension}
\end{figure}

\noindent \textbf{Hyper-parameter Selection:}
There are two critical sets of hyper-parameters in our proposed pipeline. The first one is the weights of
the primary loss ($\alpha$) and the secondary loss ($\beta$) as shown in
Equation~\ref{equation:overall_objective}. Another hyper-parameter is the dimension of HOGs. 
We estimated the values of these hyper-parameters by doing cross-validation on Validation Set.
Table~\ref{tab:ablation_loss_weights} summarises the cross-validation for weighing the contributions of the proposed losses on CaDIS validation set. We observed that setting equal contribution to the losses gives us optimal performance. We observed a similar trend on another benchmark too. This outcome also highlights the significance of the proposed auxiliary loss in our pipeline. We set the values of $\alpha$ and $\beta$ equal to 1 in the rest of the experiments.  Similarly, Figure~\ref{fig:ablation_hog_dimension} shows the performance on CaDIS Validation Set with varying the dimension of the HOGs. We can see the highest performance with the dimension of 502, which we set for the rest of the experiments. 

\begin{table}[]
    \centering
    \begin{tabular}{c|c|c }
    \hline 
    \multicolumn{2}{c|}{ Weight of losses}  & {mIOU} \\ 
     \cline{1-2}
      $\alpha$ & $\beta$ &  \\
     \hline
     0.01 & 1.0 & 81.2 \\
    %  \hline
     0.1 & 1.0 & 82.1\\
    %  \hline 
     1.0 & 1.0 & \textbf{82.3} \\
    %  \hline 
     1.0 & 0.1 & 81.7 \\
     1.0 & 0.01 & 82.0 \\
     \hline 
     \end{tabular}
    \caption{Ablation study on weights of losses.}
    \label{tab:ablation_loss_weights}
\end{table}

\begin{figure}
    \centering
    \includegraphics[trim= 0cm 0.0cm 0cm 1.5cm, width=0.95\linewidth]{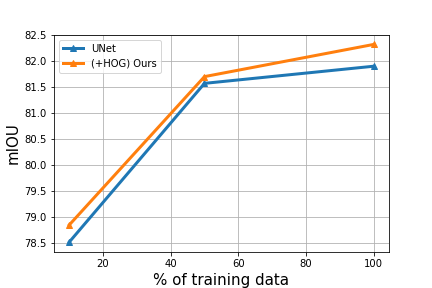}
    \caption{ A performance comparison with varying sizes of training data.}
    \label{fig:ablation_data_size}
\end{figure}

\begin{figure*}
    \centering
    \includegraphics[trim= 1cm 1cm 1cm 0cm, clip, width=0.99\textwidth]{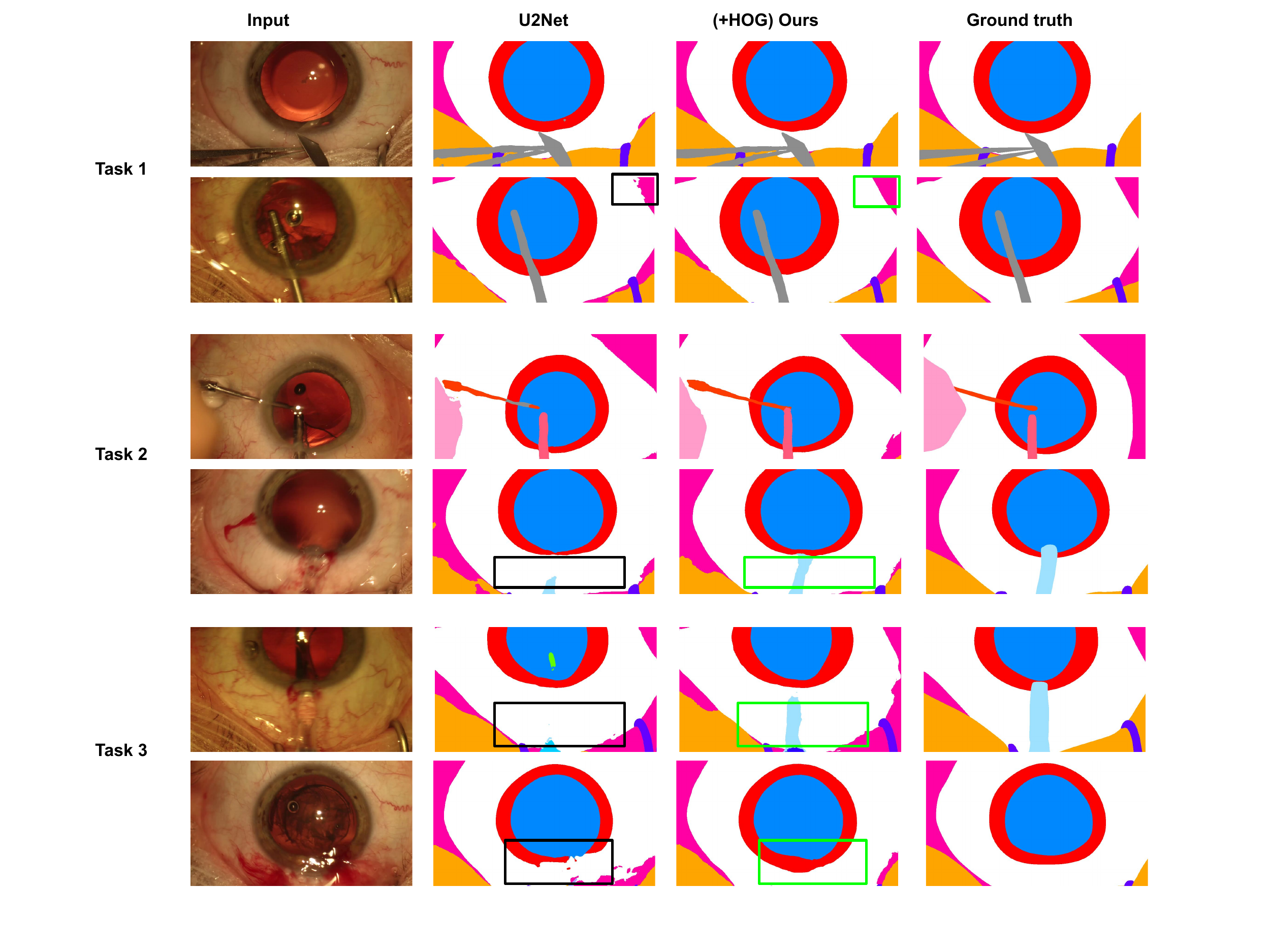}
    \caption{Qualitative comparison between the proposed method with its counter-part architecture U2Net on three different tasks. First two rows represent examples from Task 1, the middle two rows, and the last two rows are examples from Task 2 and Task 3 respectively.}
    \label{fig:qual_comparison_cataract}
\end{figure*}

\begin{figure*}
    \centering
    \includegraphics[trim= 0cm 2.5cm 0cm 1.5cm, clip, width=0.95\textwidth]{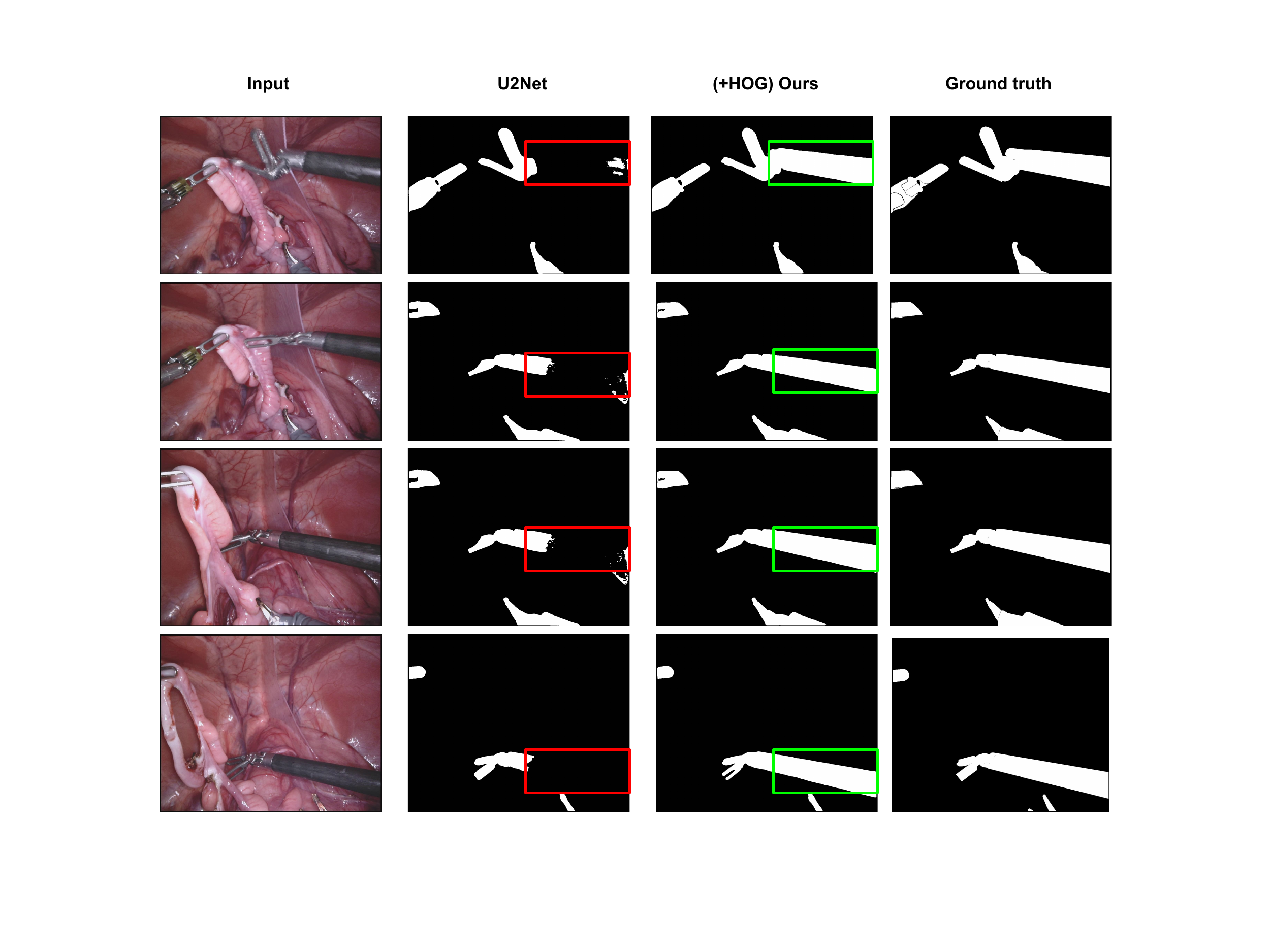}
    \caption{Qualitative comparison between before and after applying our method on U2Net in the Task 1 of robotic instrument segmentation challenge held in MICCAI 2017.}
    \label{fig:qual_instrument_seg_task1}
\end{figure*}

\begin{figure*}
    \centering
    \includegraphics[trim= 0cm 2.5cm 0cm 1.5cm, width=0.95\textwidth]{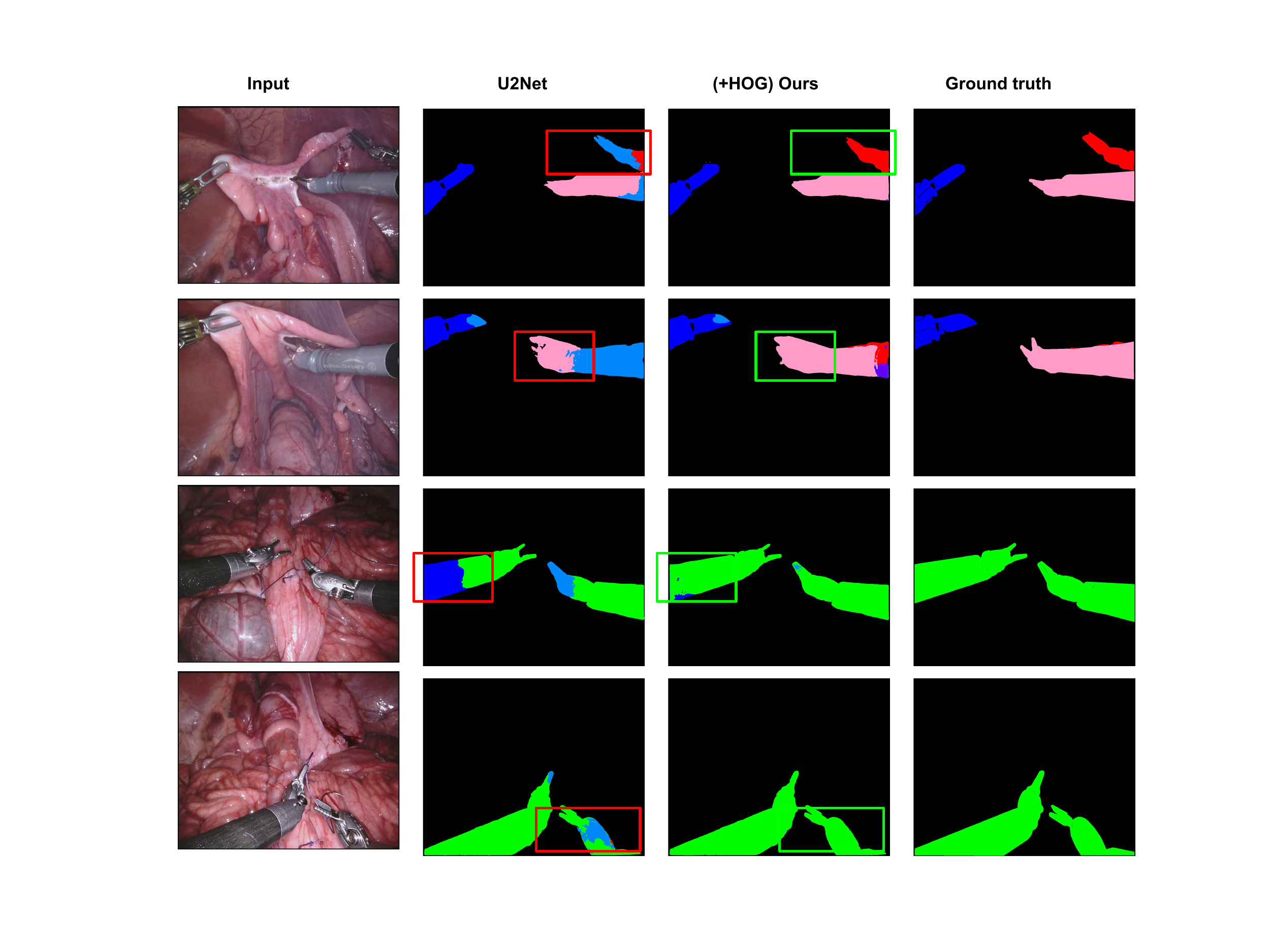}
    \caption{Qualitative comparison between before and after applying our method on U2Net in the Task 2 of robotic instrument segmentation challenge held in MICCAI 2017. }
    \label{fig:qual_instrument_seg_task2}
\end{figure*}

\begin{figure*}
    \centering
    \includegraphics[trim= 0cm 2.5cm 0cm 1.5cm, width=0.95\textwidth]{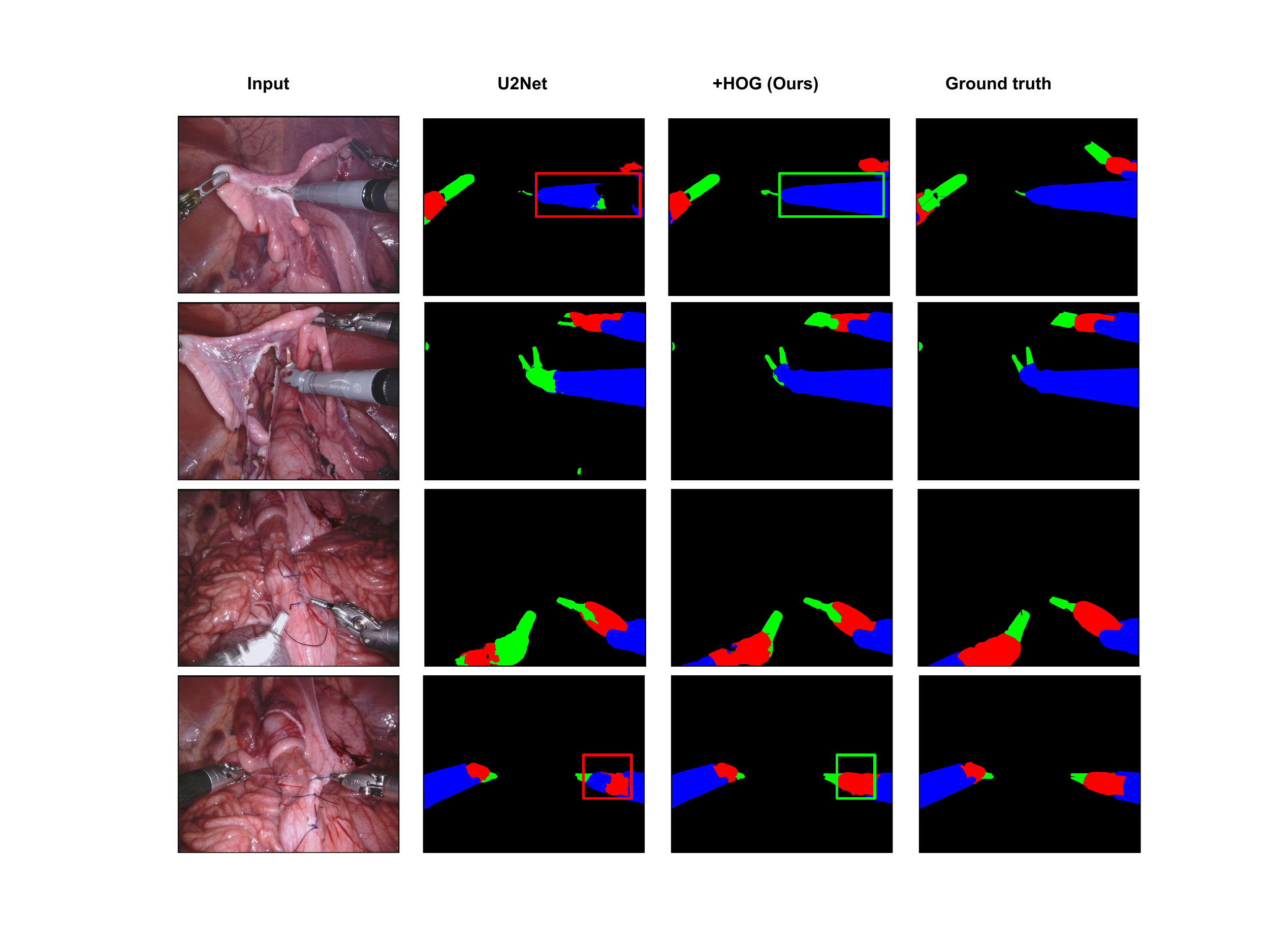}
    \caption{Qualitative comparison between before and after applying our method on U2Net in the Task 3 of robotic instrument segmentation challenge held in MICCAI 2017. }
    \label{fig:qual_instrument_seg_task3}
\end{figure*}

\noindent \textbf{Quantitative Evaluations:}
Here, we present the outcomes from our extensive experiments on two different data sets: CaDIS and Robotic Instrument Segmentation. As mentioned before, each of the benchmarks consists of three tasks resulting in six different tasks from two data sets. We extended our method on two popular baseline architectures: UNet and U2Net. We evaluate the empirical performance on the mean Intersection of Union (mIOU). 

Compared to U2Net, UNet is more efficient but is less accurate. We evaluate both the architectures on CaDIS Task 2. We choose this task due to the good trade-off of granularity and the number of training examples per category. In this experiment, UNet and U2Net obtained 81.9\% and 83.75\% mIOU, respectively. We also took a different proportion of training examples and compared the performance of UNet with/out the auxiliary task to predict HOGs. Figure~\ref{fig:ablation_data_size} summerizes our experiments. Our technique to extend UNet to a multi-task network improves the performance consistently. This gain in performance also shows that our method equally generalizes on varying sizes of training examples. For experiments on the remaining tasks from 
both the data sets, we decided to choose U2Net as our baseline architecture as its performance is clearly superior to UNet.

Table~\ref{tab:quant_cataract} summarises the performances of three different tasks on CaDIS data set. We have compared our performance with the winner of the MICCAI 2021 challenge and U2Net. From the Table, we can see that our method consistently outperforms the U2Net on both the validation set and the set. Out of 6 different scenarios, our method obtained the highest mIOU on 4 cases, slightly lagging behind the winner of MICCAI'21 challenge on Task 1. Compared to Task1, on Task 2 and Task 3, the mIOU of the winning method on MICCAI'21 dropped by a large margin (-20\%). In contrast, our cases have a slight drop in performance (-2.0\%). This shows the robustness of the proposed pipeline over the increase in granularity of the segmentation tasks. 

Similarly, Table~\ref{tab:quant_robotic_instrument} details the performance comparison on Robotic  Instrument Segmentation. We followed the evaluation protocol presented on the challenge paper and compared our performance with the winner model. In every task, our method obtained the highest mIOU surpassing the winning team's performance and our baseline U2Net by a large margin. With the increase in the granularity in the segmentation task,  the mIOU of the winner method drops by up to -50\%. At the same time, the drop in our method is only up to -30.2\%. Again, this is yet another evidence for our method being robust compared to the contemporary methods.

\noindent \textbf{Qualitative Evaluations:}
We did not limit our experiments to quantitative evaluations only. To deeper understand our method's role in improving the performance of existing architecture such as U2Net, we performed an extensive qualitative 
analysis. Figure~\ref{fig:qual_comparison_cataract} shows the qualitative comparisons of Task 1 , Task 2, and 
Task 3 on CaDIS data set. The bounding boxes locate some of the representative regions on the eye and the surgical 
instrument where U2Net fails, but our method correctly segments it. From these locations, we can see that the 
characteristics of HOGs to identify the organs and tools boundary play a crucial role in correctly segmenting the organs and the semantic parts of the surgical tools.

Similarly, Figure~\ref{fig:qual_instrument_seg_task1},
Figure~\ref{fig:qual_instrument_seg_task2}, Figure~\ref{fig:qual_instrument_seg_task3} show the qualitative 
comparison of Task 1, Task 2, and Task 3 on robotic instrument segmentation. In these qualitative analyses, we 
observe the similar trends that were seen on CaDIS data set. As we can see from these analysis, U2Net struggles quite a lot on boundary regions. Our method enables correct segmentation on such regions that we can see in our qualitative comparisons. The red bounding boxes on the Figures locates the failed cases by the baseline, whereas the green bounding boxes show the correction made by our method.

\section{Conclusions and Future Works}
In this paper, we present a novel multi-task deep learning framework for medical image segmentation. We generate the annotations of the auxiliary task in an unsupervised manner. We leverage Histogram of Oriented Gradients of images as their labels. We train the deep network jointly to minimise the losses of both the primary task, which is semantic segmentation and the auxiliary task. From our extensive qualitative and quantitative experiments on two challenging medical image segmentation benchmarks, we observe the proposed pipeline's performance superior to its counter-part single task network. In the future work, we plan to explore the higher-order statistics of hand-crafted features such as Fisher Vectors as annotation of images to train the multi-task deep semantic network.

\section{Acknowledgement}
This project is funded by the EndoMapper project by Horizon 2020 FET (GA 863146). For the purpose of open access, the author 
has applied a CC BY public copyright licence to any author accepted manuscript version arising from this submission.

% \clearpage
\bibliographystyle{splncs04}
\bibliography{refs2}
\end{document}